\documentclass[aps,prb,twocolumn,groupedaddress,showpacs]{revtex4}

\usepackage[english]{babel}
\usepackage{epsfig}

\begin{document}

\title{Thermal fluctuations in ultrasmall intrinsic Josephson junctions}

\author{A. Franz, Y. Koval, D. Vasyukov, and P. M\"uller}
\affiliation{Physikalisches Institut III, Universit\"{a}t
             Erlangen-N\"{u}rnberg, D-91058 Erlangen, Germany}
\author{H. Schneidewind}
\affiliation{Institut f\"ur Physikalische Hochtechnologie,  P.O. Box 100239, D-07702 Jena, Germany}
\author{D.A. Ryndyk\footnote{On leave from the Institute for Physics of Microstructures, RAS,
         Nizhny Novgorod, Russia}
         and J. Keller}
\affiliation{Institut f\"ur Theoretische Physik, Universit\"at
             Regensburg, D-93040 Regensburg, Germany}
\author{C. Helm}
\affiliation{Institut f\"ur Theoretische Physik, ETH Z\"urich, CH-8093
             Z\"urich, Switzerland}

\date{\today}

\begin{abstract}
Current-voltage curves of small area hysteretic intrinsic Josephson junctions, for
which the Josephson energy $E_J=\hbar J_c/2e$ is of order of thermal energy $kT$,
are investigated.  A non-monotonic temperature dependence of the switching current
is observed and explained by thermal phase fluctuations.  At low temperatures
premature switching from the superconducting into the resistive state is the most
important effect of fluctuations. At high temperatures only a single resistive
branch is observed.  At the cross-over temperature a hysteretic  phase-diffusion
branch exists. It shows the importance of a frequency-dependent impedance of an
external circuit formed by the leads.
\end{abstract}


\maketitle

\section{Introduction}

In the strongly anisotropic cuprate superconductors, such as
Bi$_2$Sr$_2$CaCu$_2$O$_{8+\delta}$ (BSCCO) or Tl$_2$Ba$_2$CaCu$_2$O$_{8+\delta}$
(TBCCO), the CuO$_2$ layers together with the intermediate material form a stack of
Josephson junctions. In the presence of a bias current perpendicular to the layers
each junction of the stack is either in the resistive or in the superconducting
state leading to the well-known multibranch structure of the IV-curves, see
Ref.\,\onlinecite{Kleiner92prl,Kleiner94prb,Mueller94inbook,Yurgens00} and
references therein.

The effect of thermal noise on the properties of Josephson junctions made from {\em
low-temperature} superconductors, and in particular on the critical current and
voltage-current curves, has been investigated experimentally and theoretically in
many publications, see
Ref.\,\onlinecite{Fulton74prb,Iansiti89prb,Kautz90prb,Martin93prb}. For {\em
high-temperature} superconductors considerable interest has been attracted to the
problems of thermal fluctuations in the vortex state, see
Ref.\,\onlinecite{Koshelev91prb,Blatter94rmp,Bulaevskii96prl,Koshelev96prl,Koshelev99prb},
but, to our knowledge, no detailed study of the influence of thermal fluctuations on
the critical current and current-voltage  curves  of mesoscopic intrinsic Josephson
junctions has been carried out. Some fluctuation effects are reported recently in
Ref.\,\onlinecite{Warburton03prb}.

\begin{figure}
  \begin{center}
  \epsfxsize=0.9\hsize
  \epsfbox{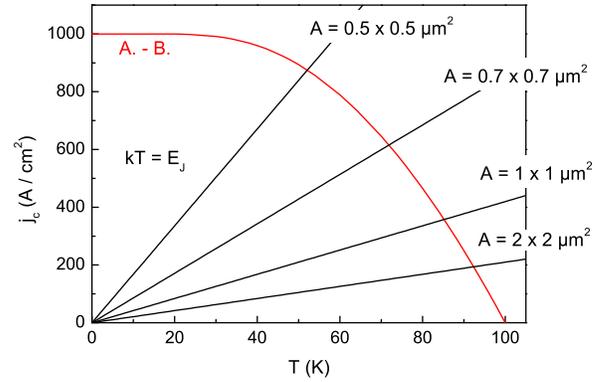}
  \caption{Estimation of crossover temperature $T^*$ from the condition $E_J(T^*)\approx kT^*$
  for mesas of different area.}
  \label{vergleich}
  \end{center}
\end{figure}

In our paper  we  present the results of experimental and theoretical investigations
of small area intrinsic Josephson junctions, when the Josephson energy
\mbox{$E_J(T)=J_c(T)/2e$} is of the order of $kT$ ($k$ is the Boltzmann constant).
For temperatures $T\ge T^*$ thermal fluctuations become important. This cross-over
temperature  can be estimated with help of Fig.\,\ref{vergleich}, where the
Ambegaokar-Baratoff temperature dependence of the critical current density $j_c(T)$
is plotted together with the linear functions $2ekT/\hbar A$ (marked as $kT=E_J$ in
the figure) for mesas with different area $A$. $T^*$ can be read-off from the
crossing points of the curves.

We discuss in the following junctions with cross-over temperatures in the range of
10-60 K \mbox{($A\!\sim 0.5\!\times\! 0.5\,\mu$m)}. Note that for all our samples
the charging energy  \mbox{$E_c= e^2/2C$} is still small compared to $E_J$, such
that quantum effects (Coulomb blockade and quantum tunneling) are not essential.
From the behaviour of the current-voltage characteristics of the junctions three
temperature regimes can be distinguished. At low temperatures, $T<T^*$, straight
lines with hysteretic jumps to other resistive branches are observed (see
Fig.\,\ref{vcc-1} and Fig.\,\ref{vcc-4}). At temperature $T\sim T^*$ the current
voltage curves are bended near the critical current (like in Fig.\,\ref{vcc-3} and
Fig.\,\ref{pd-branch}), but still some hysteresis is found. At high temperatures,
$T>T^*$, no more jumps in the current voltage curve occur and no real critical
current exists, but still a maximum in the curvature can be distinguished (see
Fig.\,\ref{vcc-2}), which can be regarded as maximum of the supercurrent. In order
to describe all temperature regimes together we introduce the concept of a
"switching" current, which will be defined below. It replaces the critical current
in the non-hysteretic high-temperature regime. A plot of the switching current as
function of temperature shows a rather non-monotonic behaviour, which is very
different from the AB-curve found for the critical current of large junctions. We
will show below that the temperature dependence of the switching current, in
particular the sharp drop,  can be explained by phase diffusion processes. This will
be done with help of an extended RSJ model containing thermal current noise and the
coupling to an external RC circuit (Fig.\,\ref{RSJ+}). The latter is necessary to
obtain the observed hysteretic behaviour for the phase diffusion branch at
intermediate temperatures $T \sim T^*$. In our case the external RC circuit is
formed most probably by a large capacitance between the ground superconducting layer
and the upper electrode together with the resistance of the contact between mesa and
upper electrode.

The paper is organized as follows. In section \ref{outline} we discuss typical
voltage-current curves at different temperatures and give a brief summary of the
theory used in calculations. In section \ref{Sample preparation} sample preparation
and experimental set-up are described. The main section \ref{results} is devoted to
experimental results and their discussion. Here also a discussion of magnetic field
effects is included.

\section{Outline of the theory}
\label{outline}

\subsection{RSJ + Johnson noise}

The simplest model to describe thermal phase fluctuations in Josephson systems is
the usual RSJ model with Johnson noise
\begin{equation}
  \frac{\hbar C}{2e}\frac{d^2\varphi}{dt^2}+\frac{\hbar}{2eR}\frac{d\varphi}{dt}
  +J_c\sin\varphi=J+J_T,
\end{equation}
\begin{equation}
  \langle J_T(t)J_T(t')\rangle=\frac{2kT}{R}\delta(t-t').
\end{equation}
Here $\varphi$ is the Josephson phase difference, $C$, $R(T)$, $J_c(T)$ are junction
capacitance, resistance, and critical current, $J$ is the bias current, $J_T(t)$ is
a random current noise. This equation can be written in dimensionless form as
\begin{equation}
  \beta\frac{d^2\varphi}{d\tau^2}+\frac{d\varphi}{d\tau}
  +\sin\varphi=j+j_T,
\end{equation}
\begin{equation}
  \langle j_T(\tau)j_T(\tau')\rangle=2\gamma\delta(\tau-\tau'),
\end{equation}
where $\tau=\omega_ct$, $j=J/J_c$, and $\omega_c=2eRJ_c/\hbar$ is the characteristic
frequency.

Here we introduce two main parameters: \\ (i) the McCumber parameter
\begin{equation}
  \beta=\frac{\omega_c^2}{\omega_p^2}=\frac{2eR^2CJ_c}{\hbar},
\end{equation}
where $\omega_p^2=2eJ_c/\hbar C$ is the Josephson plasma frequency; \\ (ii) the
fluctuation parameter (normalized temperature)
\begin{equation}
  \gamma=\frac{kT}{E_J}=\frac{2ekT}{\hbar J_c}.
\end{equation}

\subsection{Extended RSJ model }

\begin{figure}
  \begin{center}
  \epsfxsize=0.7\hsize
  \epsfbox{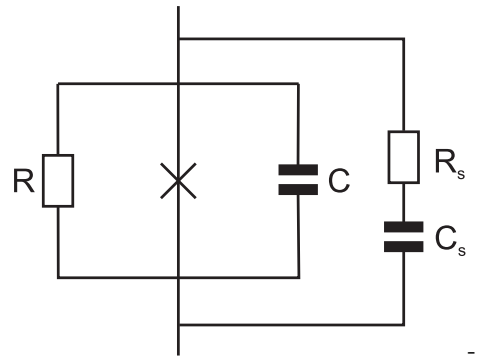}-
  \caption{RSJ model with additional frequency dependent damping modeled by
  an external RC circuit.}
  \label{RSJ+}
  \end{center}
\end{figure}

As  mentioned above, for the description of our experiments we need a frequency
dependent damping which is achieved by the coupling to an external circuit as shown
in Fig.\,\ref{RSJ+}. This model was already considered in detail by Kautz and
Martinis\cite{Kautz90prb} and is described by the following set of equations
\begin{equation}
  \beta\frac{d^2\varphi}{d\tau^2}+\frac{d\varphi}{d\tau}
  +\alpha\left(\frac{d\varphi}{d\tau}-v\right) +\sin\varphi=j+j_T+j_{Ts},
\end{equation}
\begin{equation}
  \beta\frac{dv}{d\tau}=\rho\left(\frac{d\varphi}{d\tau}-v-\frac{j_{Ts}}{\alpha}\right),
\end{equation}
where
\begin{equation}
  \langle j_{Ts}(\tau)j_{Ts}(\tau')\rangle=2\alpha\gamma\delta(\tau-\tau').
\end{equation}
Here the parameters $\alpha=R/R_s$ and $\rho=\alpha C/C_s$ are introduced, $v(t)$ is
the voltage across the external capacitance.

From the geometry of our samples we assume that $C_s$ is the capacitance between the
top and bottom contacts of the mesa and consequently $C_s\gg C$. The external
resistance $R_s$ is determined by a contact resistance and the  resistance of the
leads. For small mesas the single junction resistance $R$ is very large, up to
10-100 k$\Omega$ and $R_s\ll R$ can be assumed. Therefore for small junctions the
coupling to the external circuit should be included in the model calculations. In
the classical fluctuation regime discussed in our paper it leads to the existence of
a phase-diffusion branch instead of a zero-voltage superconducting branch.

Using this model as basis for numerical simulations we discuss in the following the
typical behaviour of current-voltage curves in the different temperature regimes. In
Section IV the model will be refined, in  order to describe the experimental results
in detail.

\subsection{Premature switching at low temperatures ($T<T^*$).}

\begin{figure}
  \begin{center}
  \epsfxsize=0.7\hsize
  \epsfbox{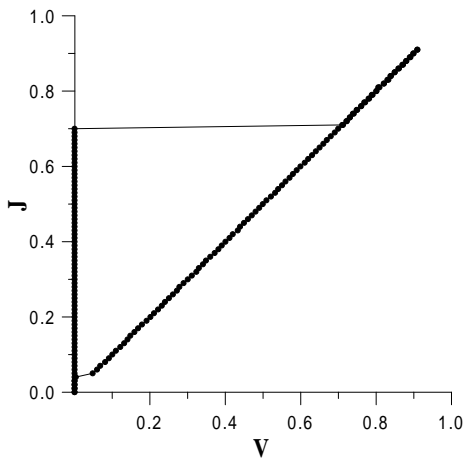}
  \caption{Typical current-voltage curves at low temperatures. The zero
  temperature critical current is suppressed by
  premature switching}
  \label{vcc-1}
  \end{center}
\end{figure}

At low temperatures a typical $I$-$V$ curve as shown in Fig.\,\ref{vcc-1} contains a
superconducting branch (S-state) with zero voltage and a hysteretic jump to the
resistive state (R-state). Due to fluctuations the critical (or, better to say
switching)  current $J_s$ is random: the first strong enough thermal kick leads to a
premature transition to the R-state, the probability of a back-transition to the
S-state is very small. The average switching current $\langle J_s\rangle(T)$ depends
on the sweeping velocity $\dot I$ of the current. It can be calculated numerically
and analytically. We found that the results of our numerical simulations are in good
agreement with the well known analytical result  $\langle J_c\rangle(T)$ at low
temperatures\cite{Tinkham96book}

\begin{equation}\label{Tinkham}
  \langle J_s\rangle(T)=J_c(T)\left[1-\left(\frac{\gamma}{2}\ln
  \frac{\omega_p\dot{I}}{2\pi I}\right)^{2/3}\right],
\end{equation}
Typically the logarithm is of the order of 10.

\subsection{Phase diffusion branch at intermediate temperatures ($T\sim T^*$)}

\begin{figure}
  \begin{center}
  \epsfxsize=0.7\hsize
  \epsfbox{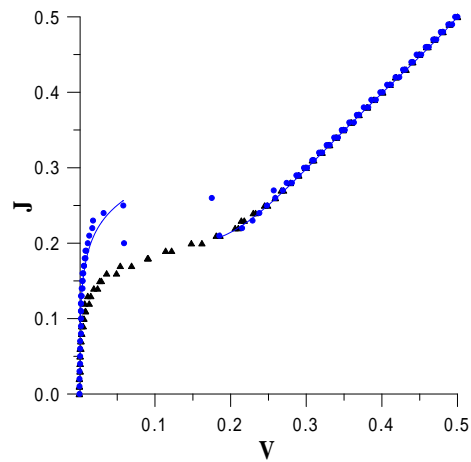}
  \caption{Typical current-voltage curves at intermediate temperatures
  with frequency-dependent damping (circles)
  and in the simple RSJ model (triangles).}
  \label{vcc-3}
  \end{center}
\end{figure}

At intermediate temperatures the S-branch shows a bending with a finite voltage
close to the switching current (see Fig.\,\ref{vcc-3}). As mentioned above this
behaviour can be explained by phase diffusion within a model with
frequency-dependent damping. Let us discuss briefly the properties of this model.

In the limit of zero frequency dissipation is determined by the junction resistance
$R$ alone, but at frequencies of the order of the plasma frequency $\omega_p$ due to
the small impedance $1/\omega_pC_s$ of the external capacitance the dissipation is
determined by the external resistance $R_s\ll R$ in parallel with $R$. Therefore for
small junctions with $R_s\ll R$ dissipation at the  plasma frequency is
significantly larger then at zero frequency. For large-$\beta$ hysteretic junctions
it leads to the coexistence of a finite-voltage phase-diffusion S-branch and
resistive branch. Thermal current-kicks stimulate jumps of the Josephson phase
difference from one potential minimum  to the next and therefore diffusion. On the
other hand, the high dissipation at the plasma frequency -- the characteristic
frequency of these jumps, prevents from a fast transition to the resistive state
with a running phase.

In Fig.\,\ref{vcc-3} the phase diffusion branch in the model with
frequency-dependent damping is shown (circles) together with the usual RSJ curve at
similar parameters (triangles).

The phase diffusion branch can be found numerically, and has been observed in our
experiments, only at intermediate temperatures. At low temperatures the voltage in
the S-state is exponentially small. At high temperatures the usual phase diffusion
takes place with a single-valued voltage-current curve.

\subsection{Switching current at high temperatures ($T>T^*$).}

\begin{figure}
  \begin{center}
  \epsfxsize=0.7\hsize
  \epsfbox{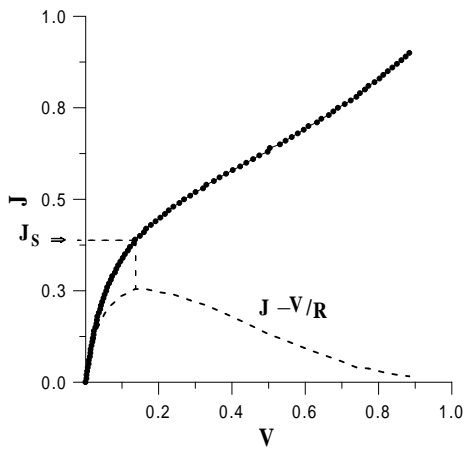}
  \caption{Typical current-voltage curves at high temperatures.
  The switching current $J_s$ is determined as  the current $J$ for
  which the contribution of an average supercurrent (dashed line) is maximal}
  \label{vcc-2}
  \end{center}
\end{figure}

At high temperatures the hysteresis disappears and only one single-valued $I$-$V$
curve with a finite voltage at all currents exists (Fig.\,\ref{vcc-2}), but the
average supercurrent, which is large  at low voltages leads to a pronounced
nonlinearity of $I$-$V$ curve. We can  extract approximately the contribution of the
supercurrent to the $I$-$V$ curve by subtracting the quasi-particle  current, which
in this model is a linear function of voltage. The current $J$ at which the
difference $J-V/R$ reaches its maximum (see Fig.\,\ref{vcc-2}) defines the switching
current $J_s$. Note that this procedure to determine a switching current can also be
applied at low temperatures  and will be generally used  later when experimental and
theoretical results are presented and compared.

\section{Sample preparation and experimental set-up}
\label{Sample preparation}

\begin{figure}
\begin{center}
    \epsfxsize=0.8\hsize
    \epsfbox{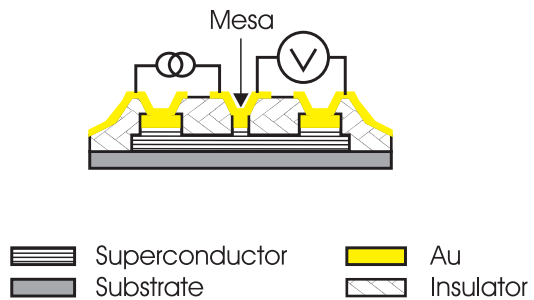}
    \caption{Sketch of the sample geometry}
    \label{schema}
\end{center}
\end{figure}

\begin{figure}
\begin{center}
    \epsfxsize=0.8\hsize
    \epsfbox{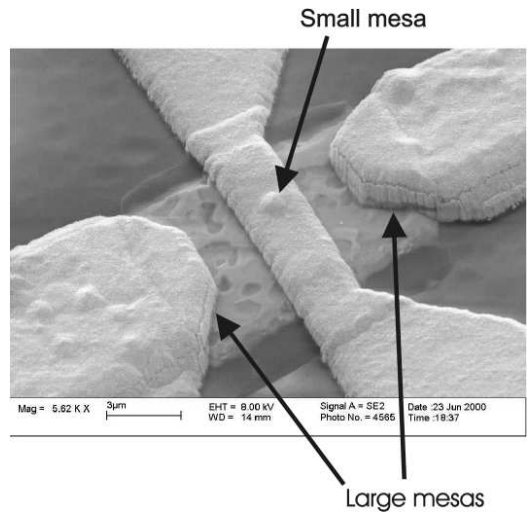}
    \caption{SEM picture of a TBCCO sample.}
    \label{emp}
\end{center}
\end{figure}

To measure the properties of ultrasmall intrinsic Josephson junctions we used a
mesa geometry as sketched in Fig.\,\ref{schema}.
Bi$_2$Sr$_2$CaCu$_2$O$_{8+\delta}$ (BSCCO) single crystals and
Tl$_2$Ba$_2$CaCu$_2$O$_{8+\delta}$ (TBCCO) thin films were used to prepare the
samples\cite{Koval00}. Fig.~\ref{emp} shows an SEM picture of a TBCCO sample. With
our technique we were able to structure samples with lateral dimensions as small as
$0.2 \times 0.2 \mu m^2$. The height of the mesas was between \mbox{60 \AA} and 1500
\AA, which is equivalent to 4 - 100 intrinsic Josephson junctions.

Several small mesas with lateral dimensions of a few $\mu m$ or lower were arranged
on one chip. On the same chip also some larger mesas with lateral dimension of
several $\mu m$ were prepared. They serve as electrical contacts to the bulk
superconductor. All mesas were contacted with Au leads. Current flows through one
large mesa into the bulk superconductor and is extracted through the small mesa or
vice versa. For voltage measurements one of the other large mesas is used. Due to
the dimensions of the small mesa it was not possible to attach two isolated Au
leads. Thus transport measurements were performed in a three-terminal configuration.

\begin{table}
\begin{tabular}{|llcccc|}
\hline sample  &  material   & width   & $J_s(4.3\,K) $ & $T_c $ & $T^*_{exp}$ \\
               &             &($\mu m$)&   $(\mu A)$    & (K)    &  (K)        \\
\hline s5 jj3  & TBCCO       & 0.7     & 16             &103     & 65          \\
CO-11-2-u-jjF  & TBCCO       & 0.5     & 12.7           &98.6    & 51          \\
CO-11-2-u-jjD  & TBCCO       & 0.5     & 5.75           & 97     & 32          \\
usm9 jj8       & BSCCO       & 0.4     & 0.74           & 86.5   & 12.5        \\
r\_el\_2 jj 20 & BSCCO       & 0.37    & 1.2            & 87.5   & 15          \\
13 r           & BSCCO       & 5       & 112            & 79.5   & -           \\
s7n4 jj1       & BSCCO       & 2       & 40             & 90     & -           \\
\hline
\end{tabular}
\caption{Sample parameters. $J_s$ is the switching current. $T^*_{exp}$ is defined
by the kink in the experimental $J_s$ vs. T plots (see Fig.\,\ref{Jc})}
\label{jj}
\end{table}

The Au electrode connected to the mesa leads to a suppression of the critical
current and critical temperature of the uppermost intrinsic Josephson
junction\cite{Kim99prb,Rother03prb}. This degeneration might be explained by the
assumption that the first superconducting layer is in close proximity with the
normal Au electrode.

One sample was prepared in a step like geometry as proposed by Wang et
al.\cite{Wang01prl}. Here we are able to record the $I$-$V$-characteristics in a
true four-point geometry (sample s7n4 jj1 Table~\ref{jj}).

The bias current was provided by a battery powered current source. The
$I$-$V$-characteristics of the samples were recorded by applying dc currents and
recording the voltages across mesas by digital voltmeters.

Measurements were done either in a standard helium dewar or in a magnet cryostat
equipped with a 5 T Helmholtz solenoid. Temperatures could be varied between 4.2~K
and room temperature. The temperature was measured with a platinum resistor for
higher temperatures and a cernox resistor for low temperatures.

To reduce external noise low pass filters were used. In the helium dewar both
cold filters and room temperature filters were used.

\section{Experimental results and discussion}
\label{results}

\subsection{Experimental voltage-current curves and $J_s(T)$}

\begin{figure}
  \begin{center}
  \epsfxsize=0.7\hsize
  \epsfbox{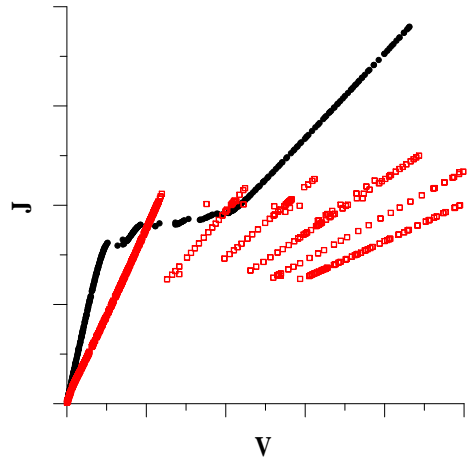}
  \caption{Typical  experimental current-voltage curves at low
  (squares) and intermediate temperatures (circles).}
  \label{vcc-4}
  \end{center}
\end{figure}

We systematically measured voltage-current curves of mesas with different
zero-temperature critical currents $J_c(0)$ at different temperatures (only a part
of the sample parameters is presented in Table.\,I). Transition from high-hysteretic
curves at low temperatures to phase-diffusion branches at crossover temperatures,
and to a single resistive curve at high temperatures was clearly observed. Examples
of experimental current-voltage  curves at low (squares) and intermediate (filled
circles) temperatures are shown in Fig.\,\ref{vcc-4} (here the contact resistance is
not subtracted). The well-known multibranch structure is observed.

\begin{figure}[t]
  \begin{center}
  \epsfxsize=0.7\hsize
  \epsfbox{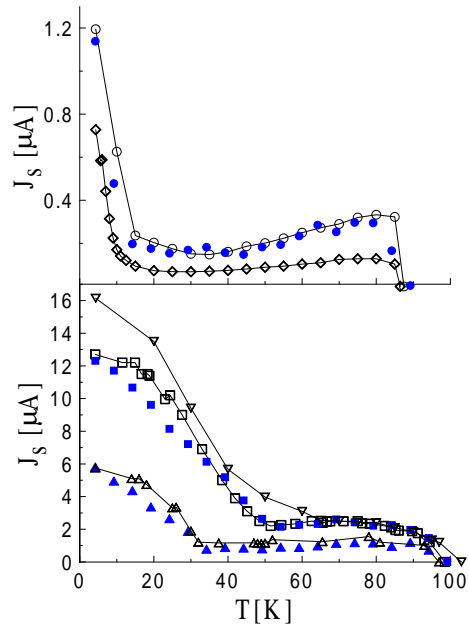}
  \caption{$J_s(T)$ for samples with different $J_c(0)$. Experiment (open symbols, the lines
  are guides to the eye),    and theory (filled symbols).}
  \label{Jc}
  \end{center}
\end{figure}

The measurements of critical  current were done mostly for the first
(superconducting) branch, such that all junctions are in the S-state at low current
and only one junction with the lowest critical current switches to the R-state. For
this branch the average switching current $J_s$ was determined. Its temperature
dependence   is shown in Fig.\,\ref{Jc} for five samples with different values of
$J_c(0)$ (open symbols) together with the results of the theoretical calculation
(filled symbols, discussed in the next section). This picture is the main result of
our work.

The main feature of all curves is a kink (or even minimum) in $J_s(T)$ at some
temperature $T^*_{exp}$. This kink is absent in the $J_s(T)$ curves of mesas with
large values of $J_c(0)$ and is shifted to lower temperatures with decreasing
$J_c(0)$. $T^*_{exp}$ corresponds approximately to the condition $E_J(T^*)=kT^*$,
which strongly suggests thermal fluctuations as an origin of the observed features.

\subsection{Theoretical voltage-current curves and $J_s(T)$}

To describe the experimental result for the switching current $J_s(T)$ in
detail it is necessary to take into account the temperature dependence of the
parameters $\beta$ and $\gamma$. These parameters are known functions of:

(i) the critical current $J_c(T)$ without fluctuations, which is determined
(qualitatively also in HTSC) by the Ambegaokar-Baratoff relation.

\begin{figure}[t]
  \begin{center}
  \epsfxsize=0.7\hsize
  \epsfbox{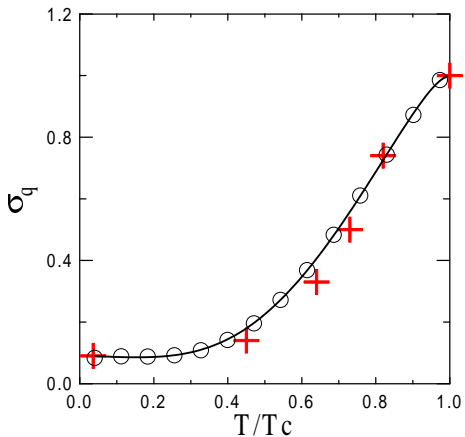}
  \caption{Quasiparticle conductivity of an intrinsic Josephson junction as
  function of $T$ shown for incoherent tunneling
  and unitary intra-layer scattering (solid line), experimental points (crosses), and
  the fit g(T) used in further calculations (circles).}
  \label{g(T)}
  \end{center}
\end{figure}

(ii) the quasiparticle resistance $R(T)$, which we determined experimentally
from current-voltage curves at small voltages (crosses in  Fig.\,\ref{g(T)}).
We also calculated $R(T)$ at $V \to 0$ from microscopic theory for different
models  of interlayer tunneling (coherent and incoherent)  and intra-layer
scattering (Born and unitary). The best fit to the experimental results was
obtained for incoherent tunneling in the unitary limit of intra-layer
scattering with an elastic scattering frequency $\nu=1$ meV. For the numerical
simulations the function $g(T)=R(0)/R(T)$ is approximated by the fit (circles
in Fig.\,\ref{g(T)}).

The results of the theoretical calculations are presented in Fig.\,\ref{Jc} for
three different zero-temperature critical currents $J_c(0)$, the
zero-temperature McCumber parameter $\beta(0)$ is in the range of $10^3$ to
$10^4$. For each temperature we calculated $J_c(T)$, $\beta(T)$, $\gamma(T)$,
and then produced current-voltage curves by numerical simulations starting from
zero external current $J$. From these results the current $J$ was determined,
at which the average supercurrent $J_{sup}=J-V/R$ reaches its maximum. This
current was defined previously as switching current $J_s$. The experimental
switching currents $J_s$ shown in Fig.\,\ref{Jc} are defined in the same way.
We find good agreement between experimental and theoretical curves. Also the
current-voltage curves calculated at different temperatures show the same
behaviour as the experimental curves.

From these results we conclude that the non-monotonic temperature dependence of
$J_s(T)$ observed in experiment can well be  explained by the model of
large-$\beta$ Josephson junction with thermal noise and  coupling to an
external impedance, if the proper temperature dependence of junction resistance
is taken into account. In particular the strong suppression of $J_s$ with
temperature at low temperatures, $T<T^*$, is due to the premature switching
effect in highly-hysteretic junctions. At high temperatures, $T>T^*$, the
McCumber parameter $\beta$ is smaller, the hysteresis disappears and the
contribution of the average supercurrent to the current at finite voltages
becomes larger, which leads to the plateau in the observed $J_s(T)$ dependence.
Further  discussions of the origins of non-monotonic temperature dependence of
$J_s$ can be found in the papers of Iansiti et al.\cite{Iansiti89prb} and Kautz
and Martinis\cite{Kautz90prb}.

\begin{figure}
  \begin{center}
  \epsfxsize=0.8\hsize
  \vskip 1cm\epsfbox{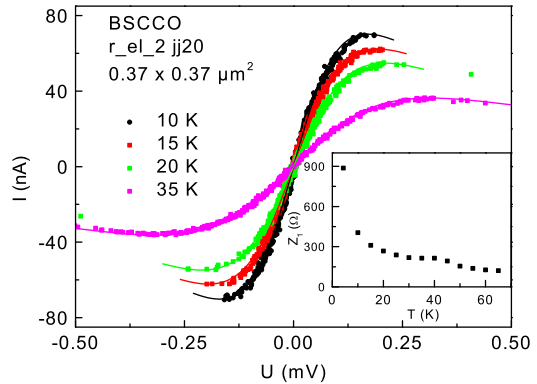}
  \caption{Phase-diffusion at low currents. Experimental points at different temperatures and
  fit to the analytical expression (\ref{pdb}). The impedance $Z_1$
  (shown in the insert) was used as a
  fitting parameter.}
  \label{pd-branch}
  \end{center}
\end{figure}

\subsection{Phase-diffusion at low currents}

For the junctions with the smallest critical current \mbox{$J_c(0)\sim 1$
$\mu$A} and highest quasiparticle resistance \mbox{$R\gg R_s$} we investigated
the low-current part of the current-voltage curves more precisely and found
clear evidence of the phase-diffusion behaviour induced by the external
circuit.

In the case $R_s\ll R$ and for $E_J<kT$ the phase-diffusion branch can be
described by the analytical expression\cite{Ingold94prb,Tinkham96book}
\begin{equation}\label{pdb}
  J(V)=\frac{4eTJ_s}{\hbar}\frac{Z_1V}{V^2+(2eZ_1T/\hbar)^2},
\end{equation}
and the small finite resistance in the S-state at very low currents and at any
temperature is given by\cite{Ingold94prb}
\begin{equation}
  R(J\rightarrow 0)=\frac{Z_1}{I_0^2(1/\gamma)-1},
\end{equation}
where $Z_1\sim R_s$ is the impedance at low frequencies (of the order of
the plasma frequency $\omega_p$) and $I_0(x)$ is a Bessel function.

The experimental points presented in Fig.\,\ref{pd-branch} can be  fitted by
Eq.\,(\ref{pdb}). The best fitting result was achieved for a temperature
dependent impedance $Z_1$ as shown in the insert. The origin of this
temperature dependence is still unclear. It may be a contribution of the
temperature dependent quasiparticle conductivity of other junctions in the
mesa: when the temperature increases this conductivity also increases and the
impedance decreases.

\subsection{$J_s(T)$ in external magnetic field}

\begin{figure}
  \begin{center}
  \epsfxsize=0.8\hsize
  \vskip 1cm\epsfbox{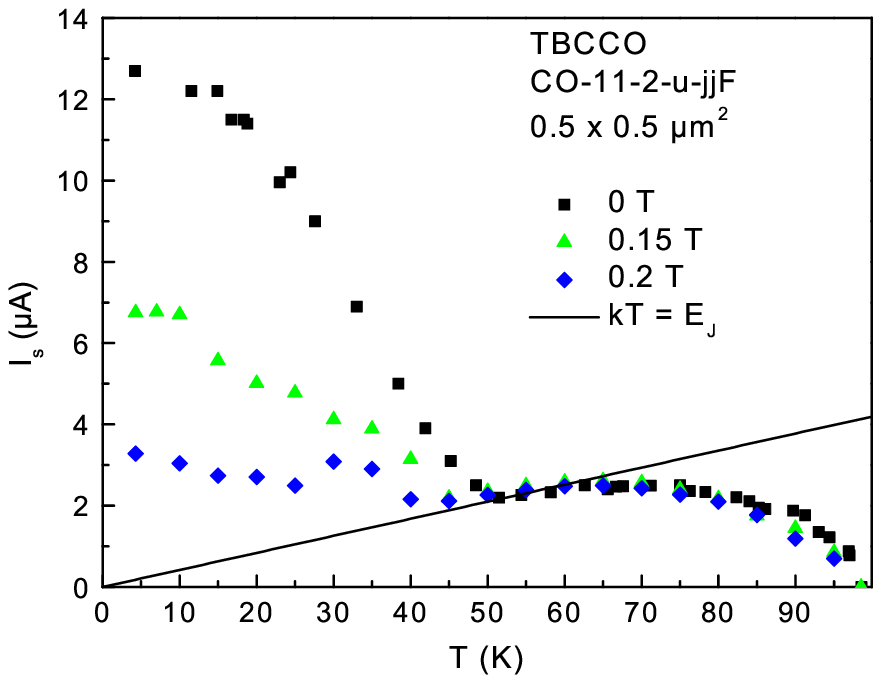}
  \epsfxsize=0.8\hsize
  \vskip 1cm\epsfbox{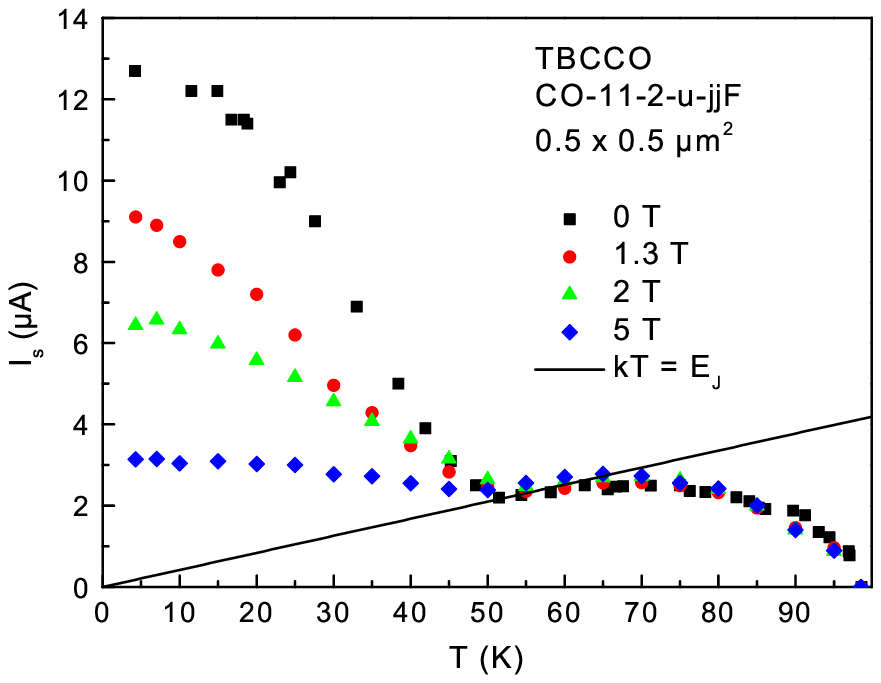}
  \caption{$J_s(T)$ at different perpendicular (upper figure) and parallel (lower figure)
  magnetic fields: experiment.}
  \label{Jc(B)}
  \end{center}
\end{figure}

Finally, we present here some results of measurements in an external magnetic
field. Magnetic fields were applied by a 5~T superconducting split coil magnet.
To measure $J_s(B)$ for parallel fields, the CuO$_2$-layers must be aligned
with high precision parallel to the magnetic field. If the magnetic field has a
component perpendicular to the layers pancake vortices enter the Josephson
junctions and suppress the critical current. The orientation of samples in a
magnetic field follows the procedure described in Ref. \onlinecite{Irie00prb}.
The junction is biased at a fixed current and a fixed magnetic field is
applied. Then the voltage change is monitored while the angle $\varphi$ between
the external field and the CuO$_2$-layers was varied with a precision rotation
stage. When measured at high enough temperatures the field-induced changes of
the $I$-$V$-characteristic were reversible and could be used for alignment.
Thus, the field orientation relative to the CuO$_2$-layers could be adjusted to
an accuracy of $0.01^\circ$.

The main and most general experimental result is presented in
Fig.\,\ref{Jc(B)}. In both perpendicular and parallel magnetic fields $J_s(T)$
{\em is not changed for $T>T^*$}, while for $T<T^*$ one finds a field dependent
suppression of $J_s(T)$.

To understand this behaviour qualitatively, we explore the well-known
sin-Gordon equation, which describes long Josephson junctions (see
Ref.\,\onlinecite{Barone,Likharev})
\begin{equation}
  \beta\frac{\partial^2\varphi}{d\tau^2}-\frac{\partial^2\varphi}{dx^2}
  +\frac{\partial\varphi}{d\tau} +\sin(\varphi+B_\parallel
  x+\delta\varphi(x,\tau))=j+j_T(x,\tau).
\end{equation}
Here the coordinate $x$ is in units of the Josephson length $\lambda_J$,
$B_\parallel$ is the component of the external magnetic field parallel to the
layers, and $\delta\varphi(x,\tau)$ is a random phase distribution produced by
pancake-vortices in perpendicular magnetic
field\cite{Bulaevskii96b,Bulaevskii97prb,Koshelev00prb}. We include the phase
variation due to the external magnetic field in the $sin$-function instead of
using the more familiar procedure, where the magnetic field is taken into
account by boundary conditions (both methods are equivalent). The random
functions $j_T(x,\tau)$ and $\delta\varphi(x,\tau)$ describing noise are
assumed to be Gaussian and $\delta$-correlated in space
\begin{equation}
  \langle j_{T}(x,\tau)j_{T}(x,\tau')\rangle=2\gamma\delta(\tau-\tau')\delta(x-x'),
\end{equation}
\begin{equation}
  \langle \delta\varphi(x,\tau)\delta\varphi(x,\tau')\rangle\propto
B_z\delta(x-x').
\end{equation}
In the last expression we accept a linear magnetic field dependence of the
correlation function and neglect the time-dependence of phase shifts assuming
that the pancakes are frozen or move slowly enough. In a more realistic
approach the space and time dependence of correlation functions of the pancake
disorder depend on a state of the vortex matter, is it liquid, glass-like, or
solid.

\begin{figure}
  \begin{center}
  \epsfxsize=0.7\hsize
  \vskip 1cm\epsfbox{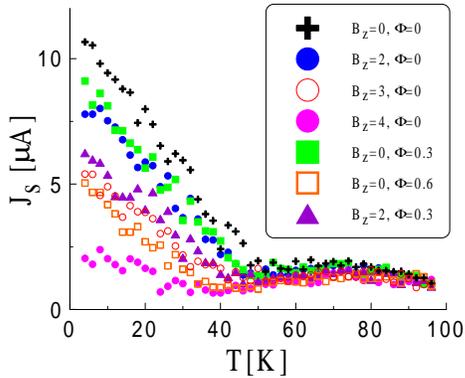}
  \caption{$J_s(T)$ at different  perpendicular (circles),
  parallel (squares), and oblique (triangles) magnetic fields: theory. $\Phi$ is the magnetic
  flux of the parallel field through a Josephson junction in units of the flux quantum $\Phi_0$.}
  \label{12(T)}
  \end{center}
\end{figure}

Results of our numerical simulations with magnetic field are presented in
Fig.\,\ref{12(T)} for several values of perpendicular and parallel fields.
The main result of our calculations (as well as similar experimental results)
is the sensitivity to a magnetic field at low temperatures and the absence of
considerable influence at high temperatures.

Qualitatively this result can be understood in the following way. The main
effect of the magnetic field is to produce phase shifts, regular for the
parallel field and random for the perpendicular field. At low temperatures,
when $J_s$ is determined by the premature switching mechanism, the reduction
of $J_c$ in the  magnetic field is important because the energy barrier becomes
smaller. But in the high-temperature regime where phase diffusion is
strong, additional phase shifts $\delta\varphi(x,\tau)$ produced by the
magnetic field cannot compete with the existing large thermal phase
fluctuations.

\section{Summary}

In this paper we discuss new experiments on the influence of thermal noise on
current-voltage  curves and the switching current of small area intrinsic
Josephson junctions. Two main effects of thermal fluctuations: premature
switching and phase-diffusion, are observed. The non-monotonic temperature
dependence of the switching current is explained theoretically with a help of
an extended RSJ model (including thermal noise and a frequency dependent
external impedance) taking into account the proper temperature dependence
of parameters.  The different behaviour of the switching current at low and
high temperatures in the presence of magnetic fields can be
explained in a similar way.

\section*{Acknowledgements}
We thank G. Blatter and L.N. Bulaevskii for valuable discussions.

This work was supported by the German Science Foundation (D.R.), and by the
Swiss National Center of Competence in Research "Materials with Novel
Electronic Properties-MaNEP" (C.H.).

\end{document}